\documentclass[
	aps,	prd,preprint,floatfix,superscriptaddress,nofootinbib
]{revtex4-1}
\usepackage{
	amsfonts,amsmath,amsthm,epsfig,amssymb,
	latexsym,eucal,array,subfigure,bm,mathrsfs,
	ulem,physics,color,graphicx,mathtools
	}

\begin{document}

\title{Global uniqueness of Kerr black hole spin and inclination from a segment of the critical curve}
\author{Kenta Hioki}
\email{kenta.hioki@gmail.com}
\noaffiliation
\begin{abstract}
High-resolution observations are expected to probe the
photon-ring structure of black hole images.
The critical curve is the geometrically defined limiting locus on the
observer's screen toward which successive higher-order photon subrings
accumulate.
For a Kerr black hole, the shape of this limiting curve depends on the
dimensionless spin parameter $a$ and the inclination angle $i$.
We investigate whether these parameters can be determined uniquely from
a connected positive-length segment of the critical curve when the
segment's position and orientation on the screen are unknown.
For rotating non-extremal Kerr black holes with $0<a<1$ and
$0<i\leq\pi/2$, we prove global uniqueness: if two such segments
coincide as point sets after orientation-preserving rigid motions of the
screen, then their Kerr parameter pairs coincide and the two rigid
motions are identical.
The proof uses an irreducible implicit polynomial to show that a
segment determines the full algebraic curve containing it and then
recovers the parameter pair $(a,i)$ from rigid-motion invariants of
that polynomial.
\end{abstract}
\maketitle

\section{Introduction}
\label{sec:intro}
Horizon-scale images of M87* and Sgr~A* obtained by the Event Horizon
Telescope have opened a new observational window on strong-field
gravity~\cite{EventHorizonTelescope:2019dse,EventHorizonTelescope:2022wkp}.
The critical curve is the geometrically defined limiting locus on the
observer's screen toward which successive higher-order photon subrings
accumulate~\cite{Bardeen:1973xx,Gralla:2019xty,Johnson:2019ljv}.
Proposed space-VLBI concepts such as the Black Hole Explorer aim to
detect and characterize the photon-ring structure at substantially
improved angular resolution~\cite{Akiyama:2024msp,Johnson:2024ttr,Lupsasca:2024xhq}.
Although future high-resolution observations may permit the critical
curve to be inferred from the accumulation of higher-order photon
subrings, only a localized segment, rather than the entire curve, may
be identifiable.
This possibility motivates the study of how much information about
black hole parameters is encoded in a segment of the critical curve.

In the idealized setup with light sources distributed uniformly at
infinity, the shadow boundary coincides with the critical curve on the
observer's sky~\cite{Hioki:2008zw,Hioki:2009na,Hioki:2023ozd}.
Accordingly, previous studies have investigated the inference of black
hole parameters from the size and shape of a shadow or critical
curve~\cite{Tsukamoto:2014tja,Abdujabbarov:2015xqa,
EventHorizonTelescope:2021dqv,Hioki:2022mdg,Tsukamoto:2024gkz}.
For Kerr--Newman black holes viewed away from the symmetry axis,
continuous degeneracies of shadow contours were excluded, while
possible discrete degeneracies remained open~\cite{Mars:2017jkk}.
Numerical analyses provided evidence against degeneracy of full shadow
contours for Kerr black holes, both for observers at spatial infinity
and at finite distances~\cite{Hioki:2009na,Hioki:2023ozd}, and for
Kerr--Newman black holes observed at spatial infinity~\cite{Hioki:2024vta}.
By contrast, a degeneracy of Reissner--Nordstr\"om shadows observed
from a finite distance was established analytically~\cite{Hioki:2024vta}.

A study introduced geometric observables for segments of the Kerr
critical curve and provided numerical evidence that the dimensionless
spin parameter $a$ and inclination angle $i$ are uniquely determined by
these observables without reconstructing the full
curve~\cite{Hioki:2026xch}.
This numerical evidence does not, however, prove that distinct
parameter pairs cannot produce congruent positive-length segments.

We formulate the problem for the two-parameter family of Kerr critical
curves modulo $\mathrm{SE}(2)$, the group of orientation-preserving
rigid motions of the screen.  For each parameter pair $(a,i)$, the
critical curve is expressed in standard Bardeen coordinates, while its
position and orientation in an image are regarded as unknown.  We ask
whether the congruence class of any connected positive-length segment
determines $(a,i)$ throughout the parameter domain.  Thus the problem
is local-to-global: the input is only a segment, while the conclusion
concerns the parameters of the full critical curve.

Our main result establishes global uniqueness for rotating
non-extremal Kerr black holes with $0<a<1$ and $0<i\leq\pi/2$.
More precisely, suppose that connected positive-length segments
corresponding to $(a,i)$ and $(a',i')$ are mapped by
orientation-preserving rigid motions $E$ and $E'$, respectively, and
that their images coincide as point sets.
We prove that $(a,i)=(a',i')$ and $E=E'$.
The result holds throughout the parameter domain above, rather than
only near a fixed parameter pair.

The proof eliminates the spherical photon-orbit radius $r_s$ from the
rational parametrization of the critical curve to obtain an
irreducible implicit polynomial $P$, while ruling out extraneous
components and repeated factors.
The irreducibility of $P$ implies that a positive-length segment
determines the full algebraic curve containing it.
An analysis of the rigid-motion symmetries of this curve identifies its
intrinsic axes and excludes nontrivial orientation-preserving
symmetries, while rigid-motion invariants of $P$ uniquely determine the
parameter pair $(a,i)$.

The limiting cases $a=0$ and $a=1$ are treated separately.
In the Schwarzschild case, the circular critical curve is independent
of $i$ and has continuous rotational symmetry.  In the extremal case,
the near-horizon line makes the inclination nonunique when arbitrary
positive-length segments are allowed.

The remainder of the paper is organized as follows.
In Sec.~\ref{sec:set}, we review Kerr null geodesics and specify the
distant observer.
In Sec.~\ref{sec:spo}, we describe the spherical photon orbits relevant
to the critical curve.
In Sec.~\ref{sec:pho}, we define segments of the critical curve.
In Sec.~\ref{sec:uni}, we develop the algebraic uniqueness argument and
discuss the limiting spin cases.
Finally, Sec.~\ref{sec:conclusion} summarizes the results and outlines
directions for future work.

Throughout this paper, we use geometrized units with $G=c=1$.
All quantities with dimensions of length are measured in units of the
black hole mass $M$, and we use the same symbols for the resulting
dimensionless quantities.

\section{Null geodesics and the distant observer}
\label{sec:set}
In this section, we review null geodesics in the Kerr spacetime and
specify the distant observer used to define the image plane.

\subsection{Kerr geometry and null-geodesic equations}
The spacetime of a rotating black hole is widely believed to be well
described by the Kerr metric~\cite{Kerr}.
In Boyer--Lindquist coordinates
$x^\mu=(t,r,\theta,\phi)$ $(\mu,\nu=0,1,2,3)$,
the line element is given by
\begin{eqnarray}
g_{\mu \nu},{\rm d}x^\mu {\rm d}x^\nu
=
-\left(
1-\frac{2r}{\varSigma}
\right) {\rm d}t^2
+
\frac{\varSigma}{\varDelta} {\rm d}r^2
+
\varSigma {\rm d}\theta ^2
-\frac{4ra\sin^2\theta}{\varSigma} {\rm d}t {\rm d}\phi
+
\frac{A \sin ^2 \theta}{\varSigma} {\rm d}\phi ^2 ,
\label{eq:metric}
\end{eqnarray}
where
\begin{eqnarray}
\varSigma
&:=&
r^2 + a^2 \cos^2 \theta, \\
\varDelta
&:=&
r^2 - 2r + a^2, \\
A
&:=&
\left( r^2 + a^2 \right)^2
- a^2 \varDelta \sin^2\theta .
\end{eqnarray}
As stated in Sec.~\ref{sec:intro}, all quantities with dimensions of
length are identified with their dimensionless counterparts.
Thus, $r$, $t$, and $a$ are dimensionless, while the angular
coordinates $\theta$ and $\phi$ are intrinsically dimensionless.
The parameter $a$ represents the dimensionless spin of the black hole.
For $|a|\leq 1$, the spacetime possesses an event horizon and describes
a Kerr black hole.
The radii of the outer and inner horizons are denoted by $r_{+}$ and
$r_{-}$, respectively.

The trajectory $x^\mu(\lambda)$ of a massless test particle representing
a light ray is governed by the null geodesic equations, where $\lambda$
is an affine parameter.
Because the Kerr spacetime possesses four independent constants of
motion in involution, its geodesic equations are completely integrable.

We begin with the geodesic Lagrangian
\begin{eqnarray}
	\mathcal{L}
	&=&
	\frac{1}{2}g_{\mu\nu}\dot{x}^\mu\dot{x}^\nu,
	\qquad
	\dot{x}^\mu
	\coloneqq
	\frac{{\rm d}x^\mu}{{\rm d}\lambda},
	\label{eq:lag}
\end{eqnarray}
where an overdot denotes differentiation with respect to $\lambda$.
Because $t$ and $\phi$ are cyclic coordinates, the conserved energy
$\mathcal{E}$ and the axial component of the angular momentum $L$ are given by
\begin{eqnarray}
	\mathcal{E}
	&\coloneqq&
	-\frac{\partial\mathcal{L}}{\partial\dot{t}}
	=
	\left(1-\frac{2r}{\varSigma}\right)\dot{t}
	+
	\frac{2ra\sin^2\theta}{\varSigma}\dot{\phi},
	\\
	L
	&\coloneqq&
	\frac{\partial\mathcal{L}}{\partial\dot{\phi}}
	=
	-\frac{2ra\sin^2\theta}{\varSigma}\dot{t}
	+
	\frac{A\sin^2\theta}{\varSigma}\dot{\phi}.
	\label{eq:angmom}
\end{eqnarray}
The Lagrangian $\mathcal{L}$ is also conserved along an affinely
parametrized geodesic.
In addition, the Kerr spacetime admits a further conserved quantity
$\mathcal{Q}$, known as the Carter constant~\cite{Chandrasekhar:1985kt}.

For null geodesics, for which $\mathcal{L}=0$, it is convenient to
introduce the dimensionless impact parameters
\begin{eqnarray}
	\xi
	\coloneqq
	\frac{L}{\mathcal{E}},
	\qquad
	\eta
	\coloneqq
	\frac{\mathcal{Q}}{\mathcal{E}^2}.
\end{eqnarray}
We normalize the affine parameter as
\begin{eqnarray}
	\tilde{\lambda}
	\coloneqq
	\mathcal{E}\lambda
\end{eqnarray}
and define the corresponding normalized four-momentum by
\begin{eqnarray}
	k^\mu
	\coloneqq
	\frac{{\rm d}x^\mu}{{\rm d}\tilde{\lambda}}.
\end{eqnarray}
With these definitions, the null geodesic equations take the following
form, which is independent of $\mathcal{E}$:
\begin{eqnarray}
	\varSigma k^t
	&=&
	\frac{A-2ra\xi}{\varDelta},
	\label{eq:velocity2}
	\\
	\varSigma k^r
	&=&
	\pm\sqrt{R},
	\label{eq:velocity0}
	\\
	\varSigma k^\theta
	&=&
	\pm\sqrt{\varTheta},
	\label{eq:velocity1}
	\\
	\varSigma k^\phi
	&=&
	\frac{2ra+\xi\csc^2\theta(\varSigma-2r)}
	{\varDelta},
	\label{eq:velocity3}
\end{eqnarray}
where
\begin{eqnarray}
	K
	&\coloneqq&
	\eta+(a-\xi)^2,
	\label{eq:cqk}
	\\
	R(r)
	&\coloneqq&
	\left(r^2+a^2-a\xi\right)^2-K\varDelta,
	\\
	\varTheta(\theta)
	&\coloneqq&
	K-\left(a\sin\theta-\xi\csc\theta\right)^2.
\end{eqnarray}

\subsection{Distant-observer frame}
We next specify the observational setup for detecting null rays.
Because our observational targets are distant black holes, we consider
an observer located sufficiently far from the black hole.
Possible choices of the observer frame include the
zero-angular-momentum-observer (ZAMO) frame~\cite{Bardeen:1973xx,Cunningham:1975zz}
and the Carter frame~\cite{Carter:1968rr}, whose observers generally
have different angular velocities in the azimuthal
direction~\cite{Chang:2020lmg}.
At sufficiently large radial distances, however, both frames
asymptotically approach the same static frame.
We therefore adopt the ZAMO frame without loss of generality in the
distant-observer limit.

We place the observer at the Boyer--Lindquist coordinates
$(r,\theta)=(r_o,i)$, where $r_o$ is the radial distance from the black
hole and $i$ is the inclination angle measured from the rotation axis.
Owing to the reflection symmetry of the Kerr spacetime across the
equatorial plane, it is sufficient to restrict the inclination angle to
\begin{eqnarray}
	i \in (0,\pi/2].
\end{eqnarray}
We exclude $i=0$ because the azimuthal coordinate and its associated
basis direction become degenerate on the rotation axis, requiring a
separate coordinate description.

For this observer, we introduce the orthonormal tetrad
\begin{eqnarray}
	e_{(t)}
	&\coloneqq&
	\sqrt{\frac{A}{\varSigma\varDelta}}
	\left(
		\partial_t
		+
		\frac{2ar}{A}\partial_\phi
	\right),
	\\
	e_{(r)}
	&\coloneqq&
	-
	\sqrt{\frac{\varDelta}{\varSigma}}
	\partial_r,
	\\
	e_{(\theta)}
	&\coloneqq&
	\frac{1}{\sqrt{\varSigma}}
	\partial_\theta,
	\\
	e_{(\phi)}
	&\coloneqq&
	-
	\sqrt{\frac{\varSigma}{A}}
	\csc\theta\,\partial_\phi .
	\label{eq:tetrad}
\end{eqnarray}
Evaluated at the observer position $(r,\theta)=(r_o,i)$, the timelike
basis vector $e_{(t)}$ represents the observer's four-velocity, while
the spacelike basis vector $e_{(r)}$ points from the observer toward the
black hole.

\section{Spherical photon orbits}
\label{sec:spo}
\subsection{Orbital-radius parametrization}
We focus on spherical photon orbits, which constitute an important class
of null geodesics relevant to photon-ring formation.
A null geodesic with a constant radial coordinate is referred to as
a spherical photon orbit.

The radius $r_s$ of a spherical photon orbit is determined by the
conditions~\cite{deVries:1999tiy}
\begin{eqnarray}
	R(r_s) = 0, \qquad
	\frac{{\rm d}R}{{\rm d}r}(r_s) = 0,
	\label{eq:spo}
\end{eqnarray}
while the polar motion must satisfy
\begin{eqnarray}
	\varTheta(\theta) \ge 0
	\label{eq:contheta}
\end{eqnarray}
along the orbit.

For $a\neq0$ and $r_s\neq1$, solving Eq.~\eqref{eq:spo} for the
conserved quantities gives~\cite{deVries:1999tiy}
\begin{eqnarray}
	\xi_s
	&=&
	\frac{r_s^2+a^2}{a}
	-
	\frac{2r_s\varDelta}{a(r_s-1)},
	\label{eq:spoell}
	\\
	\eta_s
	&=&
	-
	\frac{r_s^3\left[r_s(r_s-3)^2-4a^2\right]}
	{a^2(r_s-1)^2}.
	\label{eq:spoq}
\end{eqnarray}
These quantities are conserved along a spherical photon orbit
of radius $r_s$.
Taking into account the position of the observer,
Eqs.~\eqref{eq:contheta}, \eqref{eq:spoell}, and \eqref{eq:spoq}
imply that $r_s$ must satisfy
\begin{eqnarray}
	\left.
	\varTheta(i)
	\right|_{(\xi,\eta)=(\xi_s,\eta_s)}
	\ge 0.
\end{eqnarray}

Substituting Eqs.~\eqref{eq:spoell} and \eqref{eq:spoq}
into Eq.~\eqref{eq:cqk}, we obtain
\begin{eqnarray}
	K_s
	\coloneqq
	\left.
	K
	\right|_{(\xi,\eta)=(\xi_s,\eta_s)}
	=
	\frac{4r_s^2\varDelta}{(r_s-1)^2}.
	\label{eq:cartercnst}
\end{eqnarray}
Since $K$ is generally non-negative~\cite{Chandrasekhar:1985kt}
and we consider only spherical photon orbits outside the outer event
horizon, the radius $r_s$ must lie in
\begin{eqnarray}
	r_s \in (r_+,\infty).
	\label{eq:spoexist}
\end{eqnarray}

\subsection{Radial instability}
The condition for the radial instability of a spherical photon orbit is
\begin{eqnarray}
	\frac{{\rm d}^2R}{{\rm d}r^2}(r_s) > 0.
	\label{eq:sta}
\end{eqnarray}
When evaluated for a spherical photon orbit, the function
${\rm d}^2R/{\rm d}r^2$ has at most two real roots as a function of
$r_s$, namely $0$ and $r_u$, where
\begin{eqnarray}
	r_u
	\coloneqq
	1-\left(1-a^2\right)^{1/3}
\end{eqnarray}
is the non-negative root.
Accordingly, Eq.~\eqref{eq:sta} is equivalent to
\begin{eqnarray}
	r_s \in (r_u,\infty).
	\label{eq:spounsta}
\end{eqnarray}

Since $r_u\le r_+$ for $|a|\le 1$, combining
Eqs.~\eqref{eq:spoexist} and \eqref{eq:spounsta} yields
\begin{eqnarray}
	r_s \in (r_+,\infty).
	\label{eq:rsta}
\end{eqnarray}
Therefore, all spherical photon orbits outside the outer event horizon
that are relevant to photon-ring formation are radially unstable.

In the rotating nonextremal domain $0<a<1$ considered below, one has
$r_+>1$.  Hence Eq.~\eqref{eq:rsta} implies $r_s>r_+>1$, so neither
$a$ nor $r_s-1$ vanishes on the physical family of spherical photon
orbits.  The rational functions $\xi_s(r_s)$ and $\eta_s(r_s)$ are
therefore well defined throughout this family and provide the
orbital-radius parametrization used to construct the critical curve
and its algebraic representation in the following sections.

\section{Segments of the critical curve}
\label{sec:pho}
To formulate the uniqueness problem for a segment of the critical
curve, we first specify its representation on the observer's image
plane, or screen.
At spatial infinity, this plane is orthogonal to the line-of-sight
direction $e_{(r)}$.
We use the Bardeen coordinates $(\alpha,\beta)$ as Cartesian coordinates
on this plane~\cite{Bardeen:1973xx}.  In terms of the observer tetrad in
Eq.~\eqref{eq:tetrad}, they are
\begin{eqnarray}
\alpha
&\coloneqq&
\lim_{r_o \to \infty}
\frac{-r_o k^{(\phi)}}{k^{(t)}}
=
-\xi \csc i ,
\label{eq:alpha}
\\
\beta
&\coloneqq&
\lim_{r_o \to \infty}
\frac{r_o k^{(\theta)}}{k^{(t)}}
=
\left(
\eta
+
a^2 \cos^2 i
-
\xi^2 \cot^2 i
\right)^{1/2}.
\label{eq:beta}
\end{eqnarray}
Here
$\left(k^{(t)},k^{(r)},k^{(\theta)},k^{(\phi)}\right)$
are the tetrad components of the photon four-momentum evaluated at the
observer.
As in the preceding sections, quantities with dimensions of length are
measured in units of the black hole mass $M$; in particular,
$\alpha$, $\beta$, and $r_s$ are dimensionless.

We now take $0<a<1$ and associate each admissible spherical
photon-orbit radius $r_s$ with the screen points
\begin{eqnarray}
\gamma_\pm(r_s;a,i)
\coloneqq
\bigl(
\alpha(r_s;a,i),
\pm\beta(r_s;a,i)
\bigr),
\label{eq:defgammas}
\end{eqnarray}
where $\alpha(r_s;a,i)$ and $\beta(r_s;a,i)$ are obtained from
Eqs.~\eqref{eq:alpha} and \eqref{eq:beta} by setting
$\xi=\xi_s(r_s)$ and $\eta=\eta_s(r_s)$.
The functions $\xi_s$ and $\eta_s$ are given in
Eqs.~\eqref{eq:spoell} and \eqref{eq:spoq}.
Let $J(a,i)$ denote the range of $r_s$ corresponding to the unstable
spherical photon orbits that appear on the observer's screen.
For $0<a<1$, varying $r_s$ over $J(a,i)$ traces the critical
curve,
\begin{eqnarray}
C(a,i)
\coloneqq
\left\{
\gamma_+(r_s;a,i)
\,\middle|\,
r_s\in J(a,i)
\right\}
\cup
\left\{
\gamma_-(r_s;a,i)
\,\middle|\,
r_s\in J(a,i)
\right\}.
\label{eq:criticalcurve}
\end{eqnarray}

The object of interest in the following section is not necessarily the
entire critical curve $C(a,i)$, but a connected portion of it.
We define a segment of the critical curve to be a subset
$\mathcal{S}\subset C(a,i)$ of the form
\begin{eqnarray}
\mathcal{S}
=
\left\{
\gamma_+(r_s;a,i)
\,\middle|\,
r_s\in I_+
\right\}
\cup
\left\{
\gamma_-(r_s;a,i)
\,\middle|\,
r_s\in I_-
\right\},
\label{eq:defsegment}
\end{eqnarray}
where $I_+$ and $I_-$ are closed subintervals of $J(a,i)$, either of
which may be empty~\cite{Hioki:2026xch}.
They are chosen so that the union in Eq.~\eqref{eq:defsegment} is
connected, and at least one of them is nondegenerate.
Thus a segment is a compact connected portion of the critical curve
with positive arc length.
If an observation yields several disconnected portions of the critical
curve, each connected portion is regarded as a separate segment.

\section{Global uniqueness from segments of the critical curve}
\label{sec:uni}
We now show that a segment of the critical curve determines the Kerr
parameters globally even when its position and orientation on the screen
are unknown.
The argument treats a segment only as a point set and allows an
arbitrary rotation of the screen coordinates and an arbitrary shift of
their origin.
These transformations form $\mathrm{SE}(2)$, the group of
orientation-preserving rigid motions of the plane, and represent
changes in the choice of screen coordinates rather than a change of the
physical observer.
More precisely, the rotational part corresponds to rotating the chosen
orthonormal basis on the screen about the line of sight, with the
observer's four-velocity and line-of-sight direction held fixed, while
the translational part corresponds to shifting the coordinate origin.
Neither operation changes the inclination $i$ or the intrinsic shape of
the critical curve.
Throughout the main discussion, we restrict attention to the rotating
non-extremal domain
\begin{eqnarray}
0<a<1,
\qquad
0<i\leq\frac{\pi}{2}.
\label{eq:domain}
\end{eqnarray}
The limiting cases $a=0$ and $a=1$ are discussed separately at the end
of this section.

By global uniqueness, we mean the following statement.
Consider two connected positive-length segments associated with
parameter pairs $(a,i)$ and $(a',i')$, and let
$E,E'\in\mathrm{SE}(2)$ be orientation-preserving rigid motions of the
screen.
If the images of the two segments under $E$ and $E'$ coincide as point
sets, then $(a,i)=(a',i')$ and $E=E'$.

\subsection{Irreducible algebraic representation of the critical curve}
\label{sec:implicit}
Our purpose in this subsection is to eliminate the spherical
photon-orbit radius $r_s$ and represent the critical curve as a subset
of an irreducible real algebraic curve in the observer's image plane.
This algebraic representation will be used in the next subsection to
show that a segment determines the full curve.

The conserved quantities $\xi_s(r_s)$ and $\eta_s(r_s)$ are given as
rational functions of $r_s$ in Eqs.~\eqref{eq:spoell} and
\eqref{eq:spoq}.
For $r_s\neq1$, clearing their denominators gives the equivalent
polynomial relations
\begin{eqnarray}
g_1(r_s;\xi,a)
&\coloneqq&
r_s^3
-
3r_s^2
+
(a^2+a\xi)r_s
+
(a^2-a\xi),
\label{eq:g1}
\\
g_2(r_s;\eta,a)
&\coloneqq&
a^2\eta(r_s-1)^2
+
r_s^3
\left(
r_s^3
-
6r_s^2
+
9r_s
-
4a^2
\right).
\label{eq:g2}
\end{eqnarray}
As polynomials in $r_s$, $g_1$ and $g_2$ are monic of degrees $3$ and
$6$, respectively.
We denote their resultant~\cite{Cox:2015ode} with respect to $r_s$ by
$\mathrm{Res}\!\left(g_1,g_2,r_s\right)$.
It vanishes if and only if $g_1$ and $g_2$ have a common root in $r_s$.
Computing this resultant gives
\begin{eqnarray}
\mathrm{Res}\!\left(g_1,g_2,r_s\right)
=
4a^6(a^2-1)\Pi(\xi,\eta;a),
\label{eq:resultant}
\end{eqnarray}
where
\begin{eqnarray}
\Pi(\xi,\eta;a)
  = \sum_{k=0}^{3} c_k\,\eta^k,
\label{eq:gt-cubic}
\end{eqnarray}
with
\begin{eqnarray}
c_3 &=& a^2 - 1, \nonumber\\[2pt]
c_2 &=& 2a^2(a+\xi)^2 - 33a^2 - 3\xi^2 + 27, \nonumber\\[2pt]
c_1 &=& a^2(a+\xi)^4 + 3\,(11a^2+\xi^2)(a^2-\xi^2) + 54\,(a-\xi)^2,
\nonumber\\[2pt]
c_0 &=& (a-\xi)^3\bigl[(a+\xi)^3 + 27\,(a-\xi)\bigr].
\label{eq:gt-coeffs}
\end{eqnarray}
The polynomial $\Pi$ has degrees $6$, $3$, and $6$ in
$\xi$, $\eta$, and $a$, respectively.

We now pass to the observer's image plane.
Using Eqs.~\eqref{eq:alpha} and \eqref{eq:beta}, the screen coordinates
satisfy
\begin{eqnarray}
\xi
=
-\alpha\sin i,
\qquad
\eta
=
\beta^2
+
(\alpha^2-a^2)\cos^2 i.
\label{eq:screeninverse}
\end{eqnarray}
We therefore define
\begin{eqnarray}
P(\alpha,\beta;a,i)
\coloneqq
\Pi
\left(
-\alpha\sin i,\,
\beta^2+(\alpha^2-a^2)\cos^2 i;\,
a
\right).
\label{eq:Pdef}
\end{eqnarray}
For fixed $(a,i)$, let
\begin{eqnarray}
V(P)
\coloneqq
\left\{
(\alpha,\beta)\in\mathbb{R}^2
\,\middle|\,
P(\alpha,\beta;a,i)=0
\right\}
\label{eq:VP}
\end{eqnarray}
denote the real algebraic set defined by $P$.
To relate this algebraic set to the critical curve, let
$(\alpha,\beta)\in C(a,i)$ be arbitrary.
Then $(\alpha,\beta)=\gamma_\pm(r_s;a,i)$ for some
$r_s\in J(a,i)$ and one of the two signs.
The corresponding conserved quantities are
$(\xi,\eta)=(\xi_s(r_s),\eta_s(r_s))$.
Equations~\eqref{eq:g1} and \eqref{eq:g2} show that this $r_s$ is a
common root of $g_1$ and $g_2$ at these values of $(\xi,\eta)$.
By the defining property of the resultant, it therefore vanishes at
$(\xi,\eta)=(\xi_s(r_s),\eta_s(r_s))$.
Since $0<a<1$, the prefactor in Eq.~\eqref{eq:resultant} is nonzero,
and hence
$\Pi(\xi_s(r_s),\eta_s(r_s);a)=0$.
Moreover, Eqs.~\eqref{eq:alpha} and \eqref{eq:beta} show that the two
arguments of $\Pi$ in Eq.~\eqref{eq:Pdef} are precisely
$\xi_s(r_s)$ and $\eta_s(r_s)$.
Consequently $P(\alpha,\beta;a,i)=0$, and therefore
$(\alpha,\beta)\in V(P)$.
Since $(\alpha,\beta)\in C(a,i)$ was arbitrary, we obtain
\begin{eqnarray}
C(a,i)
\subset
V(P).
\label{eq:CsubsetVP}
\end{eqnarray}

It remains to show that $P$ is irreducible.
We do this directly in the language of algebraic curves and
elimination.
Substituting Eqs.~\eqref{eq:spoell} and \eqref{eq:spoq} into the screen
relation obtained by squaring Eq.~\eqref{eq:beta}, and clearing the
denominators, gives the polynomial equation
\begin{eqnarray}
H(r_s,\beta)
\coloneqq
a^2\sin^2 i\,(r_s-1)^2\beta^2
+
N(r_s)
=
0,
\label{eq:Hdef}
\end{eqnarray}
where
\begin{eqnarray}
N(r_s)
&\coloneqq&
r_s^6
-
6r_s^5
+
 (9+2a^2\cos^2 i)r_s^4
-
4a^2r_s^3
\nonumber
\\
&&
+
 (a^4\cos^4 i-6a^2\cos^2 i)r_s^2
+
 2a^4\cos^2 i\,(2-\cos^2 i)r_s
+
 a^4\cos^4 i.
\label{eq:Ndef}
\end{eqnarray}
For the following algebraic argument, we extend $r_s$ and $\beta$
from their physical real domains to independent complex variables.
In particular, $r_s$ is no longer restricted to the physical interval
$J(a,i)$.
We first show that $H$ is irreducible in
$\mathbb{C}[r_s,\beta]$, the ring of polynomials~\cite{Cox:2015ode} in $r_s$ and $\beta$
with complex coefficients.
At $r_s=1$,
\begin{eqnarray}
N(1)
=
4(a^2-1)(a^2\cos^2 i-1)
\neq
0
\label{eq:N1}
\end{eqnarray}
for $0<a<1$ and $0<i\leq\pi/2$.
Hence the two coefficients
$a^2\sin^2 i\,(r_s-1)^2$ and $N(r_s)$ of $H$, regarded as a polynomial in
$\beta$, have no common nonconstant factor.

Suppose, to the contrary, that $H$ is reducible.
Since $H$ has degree $2$ in $\beta$ and has no factor depending only
on $r_s$, it must factor as
\begin{eqnarray}
H
=
\left[
u(r_s)\beta+v(r_s)
\right]
\left[
w(r_s)\beta+z(r_s)
\right]
\label{eq:Hfactor}
\end{eqnarray}
with nonzero polynomials $u,v,w,z\in\mathbb{C}[r_s]$, where
$\mathbb{C}[r_s]$ denotes the ring of polynomials in $r_s$ with complex
coefficients.
Comparison of the coefficients of $\beta^2$, $\beta$, and the constant
term gives
\begin{eqnarray}
uw
&=&
a^2\sin^2 i\,(r_s-1)^2,
\label{eq:uw}
\\
uz+vw
&=&
0,
\label{eq:uzvw}
\\
vz
&=&
N.
\label{eq:vz}
\end{eqnarray}
Equation~\eqref{eq:N1} implies
$v(1)z(1)\neq0$.
If only one of $u(1)$ and $w(1)$ vanished, then
Eq.~\eqref{eq:uzvw} evaluated at $r_s=1$ would be impossible.
Thus both vanish at $r_s=1$.
Equation~\eqref{eq:uw} then requires
\begin{eqnarray}
u(r_s)
=
u_0(r_s-1),
\qquad
w(r_s)
=
w_0(r_s-1),
\end{eqnarray}
where $u_0$ and $w_0$ are nonzero constants satisfying
$u_0w_0=a^2\sin^2 i$.
Substituting these forms into Eq.~\eqref{eq:uzvw} gives
$(r_s-1)(u_0z+w_0v)=0$.
Since $\mathbb{C}[r_s]$ is an integral domain, we may cancel $r_s-1$
and obtain $u_0z+w_0v=0$.
Thus $z=-(w_0/u_0)v$, and Eq.~\eqref{eq:vz} becomes
$N=-(w_0/u_0)v^2$.
Because $N$ has degree $6$, the polynomial $v$ must have degree $3$.
Hence $N(r_s)$ is a nonzero constant multiple of the square of a cubic
polynomial.

Since we work over $\mathbb{C}$ and $N$ is monic, the nonzero constant
can be absorbed into the cubic and its leading coefficient normalized
to unity.  Thus
\begin{eqnarray}
N(r_s)
=
\left(
r_s^3
+
b r_s^2
+
c r_s
+
d
\right)^2.
\label{eq:Nsquare}
\end{eqnarray}
Comparing successively the coefficients of $r_s^5$, $r_s^4$, and
$r_s^3$ with Eq.~\eqref{eq:Ndef} yields
\begin{eqnarray}
b=-3,
\qquad
c=a^2\cos^2 i,
\qquad
d=a^2(3\cos^2 i-2).
\label{eq:bcd}
\end{eqnarray}
The coefficient of $r_s^2$ predicted by
Eq.~\eqref{eq:Nsquare} is then
\begin{eqnarray}
c^2+2bd
=
a^4\cos^4 i
-
18a^2\cos^2 i
+
12a^2,
\end{eqnarray}
whereas the actual coefficient in Eq.~\eqref{eq:Ndef} is
$a^4\cos^4 i-6a^2\cos^2 i$.
Equality would require
\begin{eqnarray}
12a^2\sin^2 i
=
0,
\end{eqnarray}
which contradicts $a>0$ and $i>0$.
Therefore $H$ is irreducible.

Let
\begin{eqnarray}
V_{\mathbb{C}}(H)
\coloneqq
\left\{
(r_s,\beta)\in\mathbb{C}^2
\,\middle|\,
H(r_s,\beta)=0
\right\}.
\end{eqnarray}
Since $H$ is irreducible, $V_{\mathbb{C}}(H)$ is an irreducible
complex algebraic curve.
Moreover, Eq.~\eqref{eq:N1} shows that this curve contains no point
with $r_s=1$.
Hence the rational function $\xi_s(r_s)$ in
Eq.~\eqref{eq:spoell} has no pole on $V_{\mathbb{C}}(H)$, and we define
\begin{eqnarray}
\Phi:\quad
V_{\mathbb{C}}(H)\ni(r_s,\beta)
&\longmapsto&
\left(
-\xi_s(r_s)\csc i,
\beta
\right)
\in\mathbb{C}^2
\label{eq:Phimap}
\end{eqnarray}
The first coordinate of $\Phi$ is regular by the preceding argument,
while the second coordinate $\beta$ is polynomial.
Hence both coordinate functions are regular on
$V_{\mathbb{C}}(H)$, and $\Phi$ is a regular algebraic map.

For the same fixed parameters $(a,i)$, define
\begin{eqnarray}
V_{\mathbb{C}}(P)
\coloneqq
\left\{
(\alpha,\beta)\in\mathbb{C}^2
\,\middle|\,
P(\alpha,\beta;a,i)=0
\right\}.
\nonumber
\end{eqnarray}
Its real locus is $V(P)$.

We now show that the image of $V_{\mathbb{C}}(H)$ under $\Phi$ is
exactly $V_{\mathbb{C}}(P)$.
We prove the two inclusions separately.

First, let $(r_s,\beta)\in V_{\mathbb{C}}(H)$.
Since $V_{\mathbb{C}}(H)$ contains no point with $r_s=1$, the
denominator-clearing step used to obtain Eq.~\eqref{eq:Hdef} from the
squared form of Eq.~\eqref{eq:beta}, after substituting
$(\xi,\eta)=(\xi_s(r_s),\eta_s(r_s))$, is reversible.
Hence $H(r_s,\beta)=0$ is equivalent to
\begin{eqnarray}
\beta^2
=
\eta_s(r_s)
+
a^2\cos^2 i
-
\xi_s(r_s)^2\cot^2 i.
\label{eq:squaredscreen}
\end{eqnarray}
Set $(\alpha,\beta)=\Phi(r_s,\beta)$.
Equation~\eqref{eq:Phimap} gives
$-\alpha\sin i=\xi_s(r_s)$.
Moreover, substituting
$\alpha=-\xi_s(r_s)\csc i$ into Eq.~\eqref{eq:squaredscreen} gives
$\beta^2+(\alpha^2-a^2)\cos^2 i=\eta_s(r_s)$.
Hence Eq.~\eqref{eq:screeninverse} yields
$(\xi,\eta)=(\xi_s(r_s),\eta_s(r_s))$.
Equations~\eqref{eq:g1} and \eqref{eq:g2} then show that $r_s$ is a
common root of $g_1$ and $g_2$, so their resultant vanishes.
Since the prefactor in Eq.~\eqref{eq:resultant} is nonzero for
$0<a<1$, it follows that
$\Pi(\xi_s(r_s),\eta_s(r_s);a)=0$.
Equation~\eqref{eq:Pdef} therefore gives
$P(\alpha,\beta;a,i)=0$, and hence
$(\alpha,\beta)\in V_{\mathbb{C}}(P)$.
Because $(r_s,\beta)\in V_{\mathbb{C}}(H)$ was arbitrary, we conclude
$\Phi\!\left(V_{\mathbb{C}}(H)\right)\subset V_{\mathbb{C}}(P)$.

Conversely, we prove
$V_{\mathbb{C}}(P)\subset
\Phi\!\left(V_{\mathbb{C}}(H)\right)$.
Let $(\alpha,\beta)\in V_{\mathbb{C}}(P)$ be arbitrary, so that
$P(\alpha,\beta;a,i)=0$.
For this complex point, define $(\xi,\eta)\in\mathbb{C}^2$ by
Eq.~\eqref{eq:screeninverse}.
Since Eq.~\eqref{eq:Pdef} defines $P$ by substituting these two
quantities into $\Pi$, the condition $P(\alpha,\beta;a,i)=0$ is
equivalent to $\Pi(\xi,\eta;a)=0$.
Equation~\eqref{eq:resultant} then gives
$\mathrm{Res}\!\left(g_1,g_2,r_s\right)=0$.
For these fixed values of $(\xi,\eta)$, Eqs.~\eqref{eq:g1} and
\eqref{eq:g2} are polynomials in the single variable $r_s$.
By the defining property of the resultant, they therefore have a common
root $r_s\in\mathbb{C}$.
This is the key reverse step: the resultant condition recovers, as a
common root, the variable $r_s$ that was eliminated in constructing
$P$.  This algebraic root need not lie in the physical interval
$J(a,i)$; such a restriction is not required for the complex-algebraic
argument.

This common root cannot equal $1$.
Indeed, Eq.~\eqref{eq:g1} gives
$g_1(1;\xi,a)=2(a^2-1)\neq0$ in the parameter range
\eqref{eq:domain}.
Since $r_s\neq1$ and $a\neq0$, the common-root equations
\eqref{eq:g1} and \eqref{eq:g2} are equivalent to the rational
relations \eqref{eq:spoell} and \eqref{eq:spoq}, respectively.
Hence $\xi=\xi_s(r_s)$ and $\eta=\eta_s(r_s)$.

Substituting these identities into Eq.~\eqref{eq:screeninverse} gives
Eq.~\eqref{eq:squaredscreen}.
Since $r_s\neq1$, Eq.~\eqref{eq:Hdef} is equivalent to
Eq.~\eqref{eq:squaredscreen}, and therefore $H(r_s,\beta)=0$.
Thus $(r_s,\beta)\in V_{\mathbb{C}}(H)$, while
Eqs.~\eqref{eq:screeninverse} and \eqref{eq:Phimap} give
$(\alpha,\beta)=\Phi(r_s,\beta)$.
Thus $(\alpha,\beta)\in
\Phi\!\left(V_{\mathbb{C}}(H)\right)$, proving the reverse inclusion.
Combining the two inclusions, we obtain
\begin{eqnarray}
V_{\mathbb{C}}(P)
=
\Phi\!\left(V_{\mathbb{C}}(H)\right),
\label{eq:VCPimage}
\end{eqnarray}
Since $V_{\mathbb{C}}(H)$ is irreducible and $\Phi$ is a regular
algebraic map, its image is irreducible.
Equation~\eqref{eq:VCPimage} therefore shows that
$V_{\mathbb{C}}(P)$ is irreducible.

The irreducibility of $V_{\mathbb{C}}(P)$ does not by itself imply that
$P$ is irreducible, because a polynomial and any positive power of it
have the same zero set.
Factor $P$ into irreducible polynomials in
$\mathbb{C}[\alpha,\beta]$, the ring of polynomials in $\alpha$ and
$\beta$ with complex coefficients.
The zero set of $P$ is the union of the zero sets of its distinct
irreducible factors, so the irreducibility of $V_{\mathbb{C}}(P)$
implies that all these factors are associates, that is, they differ
only by nonzero constants.
Consequently,
\begin{eqnarray}
P
=
\kappa F^m
\label{eq:Ppower}
\end{eqnarray}
for some irreducible
$F\in\mathbb{C}[\alpha,\beta]$, some nonzero
$\kappa\in\mathbb{C}$, and some positive integer $m$.
It remains to exclude $m>1$.

The degree-six homogeneous part of $P$ is
\begin{eqnarray}
\Phi_6(\alpha,\beta)
=
-
(\alpha^2+\beta^2)^2
\left[
(1-a^2\cos^2 i)\alpha^2
+
(1-a^2)\beta^2
\right].
\label{eq:phi6}
\end{eqnarray}
Let $F_d$ denote the highest-degree homogeneous part of $F$.
Equation~\eqref{eq:Ppower} implies that
$\Phi_6=\kappa F_d^m$.
Hence the multiplicity of every irreducible factor of $\Phi_6$ must be
divisible by $m$.

Write $\mathrm{i}=\sqrt{-1}$ for the imaginary unit, to distinguish it
from the inclination angle $i$.
Over $\mathbb{C}$, the factor $(\alpha^2+\beta^2)^2$ in
Eq.~\eqref{eq:phi6} gives the two distinct linear factors
$\alpha+\mathrm{i}\beta$ and $\alpha-\mathrm{i}\beta$, each with
multiplicity $2$.
In the parameter range \eqref{eq:domain}, the quadratic factor in
brackets has the two distinct linear factors
$\sqrt{1-a^2\cos^2 i}\,\alpha
\pm\mathrm{i}\sqrt{1-a^2}\,\beta$, each with multiplicity $1$.
Neither of these factors is proportional to
$\alpha\pm\mathrm{i}\beta$.
Indeed,
$1-a^2\cos^2 i=(1-a^2)+a^2\sin^2 i>1-a^2$, so the coefficients
multiplying $\alpha$ and $\pm\mathrm{i}\beta$ in each of these factors
are unequal, whereas they are equal in
$\alpha\pm\mathrm{i}\beta$.
Thus $\Phi_6$ has four distinct linear factors over $\mathbb{C}$:
$\alpha+\mathrm{i}\beta$ and $\alpha-\mathrm{i}\beta$ each occur with
multiplicity $2$, whereas the other two factors each occur with
multiplicity $1$.

If $m>1$, all four multiplicities would be divisible by $m$, contrary
to the simple factors in Eq.~\eqref{eq:phi6}.
Therefore $m=1$, and $P$ is irreducible in
$\mathbb{C}[\alpha,\beta]$.
It is consequently also irreducible in the real polynomial ring
$\mathbb{R}[\alpha,\beta]$, since any factorization over $\mathbb{R}$
would also be a factorization over $\mathbb{C}$.

Finally, Eq.~\eqref{eq:phi6} shows that $P$ has degree $6$ in both
$\alpha$ and $\beta$, while Eq.~\eqref{eq:Pdef} shows that $P$ depends
on $\beta$ only through $\beta^2$.

\subsection{Algebraic information contained in a segment}
\label{sec:Pmin}
We next determine how much of the polynomial $P$ is fixed by a segment
of the critical curve.
To represent the screen-coordinate freedoms described above, let
$E\in\mathrm{SE}(2)$ be an orientation-preserving rigid motion of the
screen, consisting of a rotation followed by a translation.
We identify a screen point $(\alpha,\beta)\in\mathbb{R}^2$ with the
column vector $\bm{x}\coloneqq(\alpha,\beta)^{\mathsf T}$ and write
$\bm{t}\coloneqq(t_\alpha,t_\beta)^{\mathsf T}$.
Then
\begin{eqnarray}
E(\bm{x})
=
\mathcal{R}\bm{x}
+
\bm{t},
\qquad
\mathcal{R}\in\mathrm{SO}(2).
\label{eq:rigid}
\end{eqnarray}
Let $\mathcal{S}$ be a segment of the critical curve represented as in
Eq.~\eqref{eq:defsegment}, and set $S\coloneqq E(\mathcal{S})$.
We regard $S$ as a subset of $\mathbb{R}^2$ and define its vanishing ideal~\cite{Cox:2015ode} in
$\mathbb{R}[\alpha,\beta]$ by
\begin{eqnarray}
I(S)
\coloneqq
\left\{
f\in\mathbb{R}[\alpha,\beta]
\,\middle|\,
f(p)=0
\text{ for all }p\in S
\right\}.
\label{eq:idealS}
\end{eqnarray}
The real algebraic set determined by this ideal is
\begin{eqnarray}
V\bigl(I(S)\bigr)
\coloneqq
\left\{
p\in\mathbb{R}^2
\,\middle|\,
f(p)=0
\text{ for all }f\in I(S)
\right\}.
\label{eq:varietyS}
\end{eqnarray}
Thus $V(I(S))$ is the smallest real algebraic set, in the sense of
polynomial zero sets, that contains $S$.

Since $E$ is an invertible affine transformation, the substitution
$f\mapsto f\circ E^{-1}$ is an automorphism of both
$\mathbb{R}[\alpha,\beta]$ and $\mathbb{C}[\alpha,\beta]$.
It therefore preserves irreducibility, so $P\circ E^{-1}$ is
irreducible over both $\mathbb{R}$ and $\mathbb{C}$.

The polynomial $P\circ E^{-1}$ vanishes on $S$.
Indeed, for each $p\in S$, there is a point $q$ on the original segment
such that $p=E(q)$, and hence
$(P\circ E^{-1})(p)=P(q)=0$.
Every polynomial multiple of $P\circ E^{-1}$ also vanishes on $S$.
Therefore the principal ideal~\cite{Cox:2015ode} generated by this polynomial satisfies
\begin{eqnarray}
\left\langle
P\circ E^{-1}
\right\rangle
\subset
I(S).
\label{eq:idealinclusion}
\end{eqnarray}
Here $\langle P\circ E^{-1}\rangle$ consists of all polynomial
multiples of $P\circ E^{-1}$.

We now show that the reverse inclusion also holds.
Let $f\in I(S)$ and regard it as an element of
$\mathbb{C}[\alpha,\beta]$.
Suppose that $P\circ E^{-1}$ does not divide $f$ over $\mathbb{C}$.
Since $P\circ E^{-1}$ is irreducible over $\mathbb{C}$, the two
polynomials are then coprime.
Their common zero set in $\mathbb{C}^2$ is therefore finite~\cite{Cox:2015ode}.
On the other hand, both polynomials vanish on the positive-length
segment $S$, which contains infinitely many real, and hence complex,
common points.  This is a contradiction.

Thus, for every $f\in I(S)$, there exists
$f_0\in\mathbb{C}[\alpha,\beta]$ such that
$f=(P\circ E^{-1})f_0$.
Both $f$ and $P\circ E^{-1}$ have real coefficients, so taking the
complex conjugate of the coefficients gives
$f=(P\circ E^{-1})\overline{f_0}$.
Since $\mathbb{C}[\alpha,\beta]$ is an integral domain and
$P\circ E^{-1}\neq0$, it follows that
$f_0=\overline{f_0}$.
Hence $f_0\in\mathbb{R}[\alpha,\beta]$, and therefore
$f\in\langle P\circ E^{-1}\rangle$.
This proves the reverse inclusion.  Combining the two inclusions gives
\begin{eqnarray}
I(S)
=
\left\langle
P\circ E^{-1}
\right\rangle.
\label{eq:idealgen}
\end{eqnarray}

By Eqs.~\eqref{eq:varietyS} and \eqref{eq:idealgen}, a point annihilates
all elements of $I(S)$ if and only if it annihilates the generator
$P\circ E^{-1}$.  Therefore
\begin{eqnarray}
V\bigl(I(S)\bigr)
=
V\bigl(P\circ E^{-1}\bigr).
\label{eq:varietygen}
\end{eqnarray}
Although $S$ itself is only a segment and need not be an algebraic
set, Eq.~\eqref{eq:varietygen} shows that its real algebraic closure is
the entire curve defined by $P\circ E^{-1}$.
Finally, if two polynomials generate the same principal ideal, each is
a polynomial multiple of the other.
Since $\mathbb{R}[\alpha,\beta]$ is an integral domain, the product of
these two polynomial multipliers must be $1$.
This is possible only if both multipliers are nonzero real constants.
Hence any generator of $I(S)$ differs
from $P\circ E^{-1}$ only by multiplication by a nonzero real constant.
Thus the segment determines $P\circ E^{-1}$ uniquely up to this
constant factor.

\subsection{Leading form and rigid symmetries}
\label{sec:leading}
We now extract from $P$ the geometric information needed to remove the
unknown rigid motion.
The degree-six homogeneous part $\Phi_6$ was obtained in
Eq.~\eqref{eq:phi6}.
The quadratic factor in Eq.~\eqref{eq:phi6} is represented by the
symmetric matrix
\begin{eqnarray}
\mathsf{M}(a,i)
\coloneqq
\begin{pmatrix}
1-a^2\cos^2 i & 0\\
0 & 1-a^2
\end{pmatrix}.
\label{eq:leadingmatrix}
\end{eqnarray}
Its eigenvalues are $1-a^2\cos^2 i$ and $1-a^2$, and their difference
is $a^2\sin^2 i>0$ in the parameter range \eqref{eq:domain}.
Thus $\mathsf{M}(a,i)$ has two distinct one-dimensional eigenspaces,
spanned by $(1,0)$ and $(0,1)$, respectively.
Their eigendirections are therefore the $\alpha$- and
$\beta$-directions.
The radial factor $(\alpha^2+\beta^2)^2$ in Eq.~\eqref{eq:phi6} is
invariant under every orthogonal transformation and therefore carries
no directional information.  Hence all directional dependence of the
leading form $\Phi_6$ is carried by its quadratic factor, so the
$\alpha$- and $\beta$-eigendirections identified above are precisely
the two directions distinguished by $\Phi_6$.
We call them the principal directions of the leading form.
They are intrinsic to the curve as embedded in the Euclidean screen,
up to rigid motion: translations leave $\Phi_6$ unchanged, orthogonal
changes of screen coordinates carry the two directions with them, and
multiplication of $P$ by a nonzero constant does not alter them.

Both eigenvalues in Eq.~\eqref{eq:leadingmatrix} are positive for
$0<a<1$.
Indeed, $1-a^2\cos^2 i\geq1-a^2>0$, so
Eq.~\eqref{eq:phi6} gives
\begin{eqnarray}
\Phi_6(\alpha,\beta)
\leq
-(1-a^2)(\alpha^2+\beta^2)^3.
\label{eq:phi6bound}
\end{eqnarray}
Since all remaining terms of $P$ have degree at most $5$, the negative
degree-six term dominates as $\alpha^2+\beta^2\to\infty$.
Hence $P(\alpha,\beta)\to-\infty$.
Therefore $P$ has no zeros outside a sufficiently large disk, and the
real algebraic curve $V(P)$ is bounded.

Equation~\eqref{eq:Pdef} shows that $P$ depends on $\beta$ only through
$\beta^2$.  Hence $P(\alpha,-\beta)=P(\alpha,\beta)$, and reflection
across $\beta=0$ is a symmetry of $V(P)$.

Now consider any affine reflection symmetry $\rho$ of $V(P)$, and write
$\rho(\bm{x})=Q\bm{x}+\bm{t}$, where $Q\in\mathrm{O}(2)$ is the
linear reflection across the line through the origin parallel to the
axis of $\rho$, while $\bm{t}$ accounts for the position of that axis.
For every $\bm{x}\in V(P)$, preservation of $V(P)$ gives
$P(\rho(\bm{x}))=0$.  Thus the polynomials $P$ and $P\circ\rho$ both
vanish on $V(P)$, which is infinite by Eq.~\eqref{eq:CsubsetVP}.
If $P$ did not divide $P\circ\rho$, the irreducibility of $P$ would
make these polynomials coprime, so their common zero set in
$\mathbb{C}^2$ would be finite~\cite{Cox:2015ode}, a contradiction.
Therefore $P$ divides $P\circ\rho$.
Since an invertible affine change of variables preserves degree, both
polynomials have degree $6$, and their quotient must have degree $0$.
Both have real coefficients, so
$P\circ\rho=\kappa P$ for some nonzero real constant $\kappa$.

We now compare the degree-six homogeneous parts of this identity.
The degree-six part of $P\circ\rho$ is $\Phi_6(Q\bm{x})$: every term
involving the translation $\bm{t}$ has degree at most $5$, and the
terms of $P$ below degree $6$ cannot produce a degree-six term under
affine substitution.  Therefore
$\Phi_6(Q\bm{x})=\kappa\Phi_6(\bm{x})$.

The radial factor $(\alpha^2+\beta^2)^2$ in
Eq.~\eqref{eq:phi6} is invariant under $Q$.
Writing the remaining quadratic factor as
$\bm{x}^{\mathsf T}\mathsf{M}(a,i)\bm{x}$, the preceding identity
therefore gives
\begin{eqnarray}
Q^{\mathsf T}\mathsf{M}(a,i)Q
&=&
\kappa\mathsf{M}(a,i).
\label{eq:Mcongruence}
\end{eqnarray}
Taking the trace of Eq.~\eqref{eq:Mcongruence} and using the
orthogonality of $Q$ gives
$\operatorname{tr}\mathsf{M}
=\kappa\operatorname{tr}\mathsf{M}$.
Since both eigenvalues of $\mathsf{M}(a,i)$ are positive,
$\operatorname{tr}\mathsf{M}>0$, and hence $\kappa=1$.
Equation~\eqref{eq:Mcongruence} is then equivalent to
$\mathsf{M}Q=Q\mathsf{M}$, so $Q$ preserves each eigenspace of
$\mathsf{M}$.
Because the eigenvalues of $\mathsf{M}$ are distinct, these eigenspaces
are precisely the one-dimensional $\alpha$- and $\beta$-directions.

For the linear part $Q$ of an affine reflection $\rho$, vectors
parallel to the reflection axis are fixed, whereas vectors perpendicular
to it change sign.
Thus the $+1$ eigenspace of $Q$ is the direction parallel to the
reflection axis, while its $-1$ eigenspace is the perpendicular
direction.
Since $Q$ preserves the $\alpha$- and $\beta$-directions, its $+1$
eigenspace must be one of these two directions.
Therefore the reflection axis of $\rho$ must be parallel to either the
$\alpha$- or the $\beta$-axis.

There cannot be a second reflection axis parallel to the
$\alpha$-axis.
Indeed, suppose that $\beta=b$, with $b\neq0$, were such an axis.
Let $\rho_0(\alpha,\beta)=(\alpha,-\beta)$ and
$\rho_b(\alpha,\beta)=(\alpha,2b-\beta)$ denote the reflections across
$\beta=0$ and $\beta=b$, respectively.  Their composition is
\begin{eqnarray}
(\rho_b\circ\rho_0)(\alpha,\beta)
=
(\alpha,\beta+2b),
\label{eq:parallelreflections}
\end{eqnarray}
which is a nonzero translation.
If both reflections preserved $V(P)$, this translation would also
preserve it.
Hence $(\alpha,\beta+2nb)\in V(P)$ for every
$(\alpha,\beta)\in V(P)$ and every $n\in\mathbb{Z}$.
Since $b\neq0$, these points are unbounded as $|n|\to\infty$, which
contradicts the boundedness of $V(P)$ proved above.

It remains to exclude a reflection axis parallel to the $\beta$-axis.
For the remainder of the proof, we use the squared parameters
\begin{eqnarray}
A\coloneqq a^2,
\qquad
C\coloneqq\cos^2 i.
\end{eqnarray}
The physical parameter range $0<a<1$ and
$0<i\leq\pi/2$ is then equivalent to
$0<A<1$ and $0\leq C<1$.
These variables simplify both the following symmetry argument and the
parameter-recovery argument below.
The coefficient of $\beta^4$ in $P$ is the quadratic polynomial
\begin{eqnarray}
q_4(\alpha)
&=&
\left[A(C+2)-3\right]\alpha^2
-
4a^3\sin i\,\alpha
\nonumber
\\
&&
-3A^2C
+
2A^2
+
3AC
-
33A
+
27.
\label{eq:q4}
\end{eqnarray}
Suppose that the curve were symmetric about a line $\alpha=h$.
Then $P(2h-\alpha,\beta)$ vanishes on the same curve as $P$.
Since $P$ is irreducible, these two polynomials differ by a nonzero
constant factor; that is,
$P(2h-\alpha,\beta)=\kappa P(\alpha,\beta)$ for some
$\kappa\in\mathbb{R}\setminus\{0\}$.
The coefficient of $\beta^6$ on the left-hand side is $A-1$, whereas
that on the right-hand side is $\kappa(A-1)$.
Since $A-1\neq0$, we have $\kappa=1$, and hence
\begin{eqnarray}
P(2h-\alpha,\beta)
=
P(\alpha,\beta).
\end{eqnarray}
Comparing the coefficients of $\beta^4$ gives
$q_4(2h-\alpha)=q_4(\alpha)$, so that $h$ must be the symmetry axis of
the quadratic $q_4$.
Moreover, $A(C+2)-3<0$ in the parameter range under consideration, so
the quadratic coefficient in Eq.~\eqref{eq:q4} is nonzero.
The usual formula for the axis of a quadratic therefore gives
\begin{eqnarray}
h
=
h_4
\coloneqq
\frac{2a^3\sin i}{A(C+2)-3}.
\label{eq:h4}
\end{eqnarray}
On the other hand, let $p_0(\alpha)\coloneqq P(\alpha,0)$.
Its coefficients of $\alpha^6$ and $\alpha^5$ are
\begin{eqnarray}
c_{6,0}
=
AC-1,
\qquad
c_{5,0}
=
-4a^3C\sin i.
\end{eqnarray}
Symmetry about $\alpha=h$ would require
$p_0(h+x)=p_0(h-x)$, so that $p_0(h+x)$ is an even polynomial in $x$.
In particular, its coefficient of $x^5$, namely
$6c_{6,0}h+c_{5,0}$, must vanish.
Thus the same reflection axis would instead have to satisfy
\begin{eqnarray}
h
=
h_0
\coloneqq
-\frac{c_{5,0}}{6c_{6,0}}
=
\frac{2a^3C\sin i}{3(AC-1)}.
\label{eq:h0}
\end{eqnarray}
Their difference is
\begin{eqnarray}
h_4-h_0
=
-\frac{
2a^3\sin i\,
(C-1)(AC-3)
}{
3(AC-1)\left[A(C+2)-3\right]
},
\label{eq:hdiff}
\end{eqnarray}
Here $a^3\sin i>0$, $C-1<0$, and $AC-3<0$.
The denominator is also nonzero because $AC-1<0$ and
$A(C+2)-3<0$.
Consequently, Eq.~\eqref{eq:hdiff} is nonzero throughout
$0<A<1$ and $0\leq C<1$, and hence $h_4\neq h_0$.
Thus no single line $\alpha=h$ can satisfy both necessary symmetry
conditions.
No reflection axis parallel to the $\beta$-axis can therefore exist,
and the $\alpha$-axis is the unique reflection axis of $V(P)$.

We finally exclude nontrivial orientation-preserving rigid-motion
symmetries.
Every orientation-preserving rigid motion of the plane is either a
translation or a rotation about a point.
Since $V(P)$ is bounded, it cannot be invariant under a nonzero
translation: iterating such a translation would produce an unbounded
sequence of points on the curve.
Now consider a rotational symmetry and let $U\in\mathrm{SO}(2)$
denote its linear part.
The same irreducibility and degree argument used above for reflections
shows that the defining polynomial is preserved up to a nonzero real
constant.  Comparing the degree-six homogeneous parts, to which the
location of the rotation center contributes only terms of degree at
most $5$, and repeating the trace argument following
Eq.~\eqref{eq:Mcongruence}, we obtain
\begin{eqnarray}
\Phi_6(U\bm{x})
=
\Phi_6(\bm{x}),
\qquad
U^{\mathsf T}\mathsf{M}U
=
\mathsf{M}.
\end{eqnarray}
Thus $U$ preserves each eigenspace of $\mathsf{M}$.
Since the two eigenvalues in Eq.~\eqref{eq:leadingmatrix} are distinct,
these eigenspaces are the $\alpha$- and $\beta$-directions.
A plane rotation preserving both directions has angle either $0$ or
$\pi$.  The first case is the identity, so the only remaining
nontrivial possibility is a half-turn.

Suppose that a half-turn were a symmetry, and let $L$ denote the
$\alpha$-axis and $\sigma_L$ the reflection across $L$.
Conjugating $\sigma_L$ by the half-turn gives the reflection across the
image of $L$ under the half-turn.
Since $L$ is the unique reflection axis, this image must equal $L$,
which forces the center of the half-turn to lie on $L$.
The composition of the half-turn with $\sigma_L$ is then the reflection
across the line through its center perpendicular to $L$.
This would give a second reflection axis, contradicting the uniqueness
of $L$.
Hence no half-turn symmetry exists, and the identity is the only
orientation-preserving rigid-motion symmetry of the curve.

\subsection{Rigid-motion invariants and global uniqueness}
\label{sec:main}
We now combine the preceding results to prove the global-uniqueness
statement formulated at the beginning of this section.
The argument has four steps: the observed segment determines the
transformed curve polynomial up to scale; the leading form and the
unique reflection axis remove the unknown rotation; two scalar
invariants recover the parameters; and the absence of a nontrivial
orientation-preserving symmetry then fixes the remaining rigid motion.
Let
$\mathcal{S}\subset C(a,i)$ and
$\mathcal{S}'\subset C(a',i')$
be segments of the critical curves represented as in
Eq.~\eqref{eq:defsegment}, where both parameter pairs lie in the domain
\eqref{eq:domain}.
For $E,E'\in\mathrm{SE}(2)$, set
$S\coloneqq E(\mathcal{S})$ and
$S'\coloneqq E'(\mathcal{S}')$.
Suppose that their images coincide as point sets:
\begin{eqnarray}
S=S'
\label{eq:Sequal}
\end{eqnarray}
We show that this implies
\begin{eqnarray}
(a,i)
=
(a',i'),
\qquad
E
=
E'.
\label{eq:mainunique}
\end{eqnarray}
Thus a segment determines $(a,i)$ globally, independently of an
unknown rotation and translation of the screen.

The equality of the point sets in Eq.~\eqref{eq:Sequal} gives
$I(S)=I(S')$.
Applying Eq.~\eqref{eq:idealgen} to both segments shows that this common
ideal is generated both by $P\circ E^{-1}$ and by
$P\circ E'^{-1}$, with their respective parameter pairs.
Two generators of the same principal ideal differ by a unit of
$\mathbb{R}[\alpha,\beta]$, namely a nonzero real constant.
Therefore
\begin{eqnarray}
(P\circ E^{-1})(\alpha,\beta;a,i)
=
\kappa
(P\circ E'^{-1})(\alpha,\beta;a',i'),
\qquad
\kappa\in\mathbb{R}\setminus\{0\}.
\label{eq:rigidprop}
\end{eqnarray}
Thus the common segment determines the entire transformed curve
polynomial, although its normalization remains undetermined.

The degree-six leading form \eqref{eq:phi6} is unaffected by
translations and determines two orthogonal principal directions, namely
the eigendirections of its nonradial quadratic factor.
For a rigid-motion transform of $P$, these are the images of the
$\alpha$- and $\beta$-directions.
The leading form identifies the two directions but, by itself, does not
assign them their geometric roles.
The unique reflection axis established above supplies this distinction:
it is the image of the $\alpha$-axis, while the other principal direction
is the image of the $\beta$-direction.
Since the common segment determines the same transformed curve, this
reflection axis is common to the two polynomials in
Eq.~\eqref{eq:rigidprop}.

Choose Cartesian coordinates $(x,y)$ whose $x$-axis is this common
reflection axis, so that the axis is given by $y=0$ and the $y$-direction
is perpendicular to it.
This choice removes the arbitrary rotation and the translation normal
to the reflection axis.
Only the position of the origin along the axis and the choice of the
positive $x$-direction remain arbitrary.
The sign of the $y$-coordinate is immaterial because reflection across
$y=0$ is a symmetry of the curve.
Thus the remaining coordinate freedom can be represented by
$x\mapsto\varepsilon x+\tau$, where $\varepsilon=\pm1$.

Let $P_S(x,y)$ be a representative of the common polynomial in
Eq.~\eqref{eq:rigidprop}.
Equivalently, $P_S$ is a generator of the principal ideal determined by
the common segment.
Such a generator is unique only up to multiplication by a nonzero real
constant, so the normalization of $P_S$ remains arbitrary.
Write
\begin{eqnarray}
P_S(x,y)
=
\sum_{j+k\leq6}c_{j,k}x^jy^k,
\nonumber
\end{eqnarray}
where $c_{j,k}$ denotes the coefficient of $x^jy^k$.
The coefficient of $y^4$ is the quadratic polynomial
\begin{eqnarray}
q_4(x)
=
c_{2,4}x^2+c_{1,4}x+c_{0,4},
\nonumber
\end{eqnarray}
so that $P_S$ contains the term $q_4(x)y^4$.
We define the following two combinations of its coefficients:
\begin{eqnarray}
\mathcal{K}_1
&\coloneqq&
\frac{c_{6,0}}{c_{0,6}},
\label{eq:K1}
\\
\mathcal{K}_2
&\coloneqq&
\frac{\Delta_4}{c_{0,6}^2},
\qquad
\Delta_4
\coloneqq
\operatorname{disc} q_4
=
c_{1,4}^2-4c_{2,4}c_{0,4}.
\label{eq:K2}
\end{eqnarray}

We now verify that these combinations are independent of the remaining
choices.
Under the residual coordinate change, $P_S$ is replaced by
$P_S(\varepsilon x+\tau,y)$, where $\varepsilon=\pm1$.
The coefficient of $x^6$ remains $c_{6,0}$ because
$\varepsilon^6=1$, and no term of lower degree in $x$ can produce an
$x^6$ term.
The coefficient $c_{0,6}$ is also unchanged because the $y$-coordinate
is unchanged.
Moreover, the coefficient of $y^4$ becomes
$q_4(\varepsilon x+\tau)$, whose coefficients are
\begin{eqnarray}
c_{2,4},
\qquad
\varepsilon(2c_{2,4}\tau+c_{1,4}),
\qquad
c_{2,4}\tau^2+c_{1,4}\tau+c_{0,4},
\nonumber
\end{eqnarray}
and hence its discriminant is
\begin{eqnarray}
\left[\varepsilon(2c_{2,4}\tau+c_{1,4})\right]^2
-
4c_{2,4}
\left(c_{2,4}\tau^2+c_{1,4}\tau+c_{0,4}\right)
=
c_{1,4}^2-4c_{2,4}c_{0,4}.
\nonumber
\end{eqnarray}
Thus $\Delta_4$, as well as $c_{6,0}$ and $c_{0,6}$, is unchanged by
the residual coordinate freedom.

It remains to account for the arbitrary normalization of $P_S$.
If $P_S$ is multiplied by a nonzero constant $\kappa$, then
$c_{6,0}$ and $c_{0,6}$ are multiplied by $\kappa$.
The polynomial $q_4$ is also multiplied by $\kappa$, so its
discriminant $\Delta_4$ is multiplied by $\kappa^2$.
Therefore the factor $\kappa$ cancels in $\mathcal{K}_1$, while the
factor $\kappa^2$ cancels between the numerator and denominator of
$\mathcal{K}_2$.
Consequently, $\mathcal{K}_1$ and $\mathcal{K}_2$ are determined by the
segment without knowing its original position, orientation, or the
normalization of its defining polynomial.

Because the two combinations are invariant under the coordinate and
normalization freedoms just discussed, their values may be evaluated
using the standard polynomial $P$ in the Bardeen coordinates
$(x,y)=(\alpha,\beta)$.
The leading form in Eq.~\eqref{eq:phi6} gives the pure degree-six
coefficients $c_{0,6}$ and $c_{6,0}$, while the constant term of
$q_4$ in Eq.~\eqref{eq:q4} gives $c_{0,4}$.
Explicitly,
\begin{eqnarray}
c_{0,6}
&=&
A-1,
\nonumber
\\
c_{6,0}
&=&
AC-1,
\nonumber
\\
c_{0,4}
&=&
-3A^2C
+
2A^2
+
3AC
-
33A
+
27.
\label{eq:coeffs}
\end{eqnarray}
Equation~\eqref{eq:q4} also gives
$c_{2,4}=A(C+2)-3$ and $c_{1,4}=-4a^3\sin i$.
Since $A=a^2$ and $C=\cos^2 i$, we have
$c_{1,4}^2=16A^3(1-C)$.
Substitution into
$\Delta_4=c_{1,4}^2-4c_{2,4}c_{0,4}$ therefore gives
\begin{eqnarray}
\Delta_4
=
c_{1,4}^2
-
4c_{2,4}c_{0,4}
=
12(A-1)
\left(
A^2C^2
+
6AC
+
24A
-
27
\right).
\label{eq:Delta4}
\end{eqnarray}
Because $0<A<1$, the coefficient $c_{0,6}=A-1$ is nonzero.
Substitution into Eqs.~\eqref{eq:K1} and \eqref{eq:K2} consequently
gives
\begin{eqnarray}
\mathcal{K}_1
=
\frac{1-AC}{1-A},
\label{eq:K1AC}
\end{eqnarray}
and
\begin{eqnarray}
\mathcal{K}_2
=
12
\frac{
A^2C^2+6AC+24A-27
}{
A-1
}.
\label{eq:K2AC}
\end{eqnarray}
The first invariant is determined entirely by the leading form, while
the second adds information from the coefficient of $y^4$.

Equation~\eqref{eq:rigidprop} states that the polynomials associated
with the common segment differ only by a rigid change of coordinates
and a nonzero scalar factor.
The invariance proved above therefore implies
\begin{eqnarray}
\mathcal{K}_1(A,C)
=
\mathcal{K}_1(A',C'),
\qquad
\mathcal{K}_2(A,C)
=
\mathcal{K}_2(A',C').
\nonumber
\end{eqnarray}
To conclude that the parameter pairs agree, it remains to prove that
the map $(A,C)\mapsto(\mathcal{K}_1,\mathcal{K}_2)$ is injective on
$0<A<1$ and $0\leq C<1$.

To prove this injectivity over the entire parameter domain, let
\begin{eqnarray}
X
\coloneqq
1-A.
\end{eqnarray}
First, Eq.~\eqref{eq:K1AC} gives
$\mathcal{K}_1-1=A(1-C)/(1-A)>0$.
Moreover, $X>0$, and solving the same equation for $AC$ gives
$AC=1-\mathcal{K}_1X$.
Since $AC\geq0$, the admissible range of $X$ is therefore
\begin{eqnarray}
\mathcal{K}_1>1,
\qquad
0<X\leq\frac{1}{\mathcal{K}_1},
\qquad
AC
=
1-\mathcal{K}_1X.
\label{eq:Xrange}
\end{eqnarray}
Substitution into Eq.~\eqref{eq:K2AC} yields
\begin{eqnarray}
\frac{\mathcal{K}_2}{12}
=
8\mathcal{K}_1
+
24
-
\frac{4}{X}
-
\mathcal{K}_1^2X.
\label{eq:K2X}
\end{eqnarray}
For fixed $\mathcal{K}_1$, the derivative of the right-hand side is
\begin{eqnarray}
\frac{4}{X^2}
-
\mathcal{K}_1^2
\geq
3\mathcal{K}_1^2
>
0
\end{eqnarray}
throughout the interval in Eq.~\eqref{eq:Xrange}.
Thus the right-hand side of Eq.~\eqref{eq:K2X} is strictly increasing
throughout the full admissible interval, and the equation has at most
one admissible solution for $X$.
Because the invariant values under consideration arise from an
admissible parameter pair $(A,C)$, the value $X=1-A$ provides an
admissible solution of Eq.~\eqref{eq:K2X}.
Together with the preceding at-most-one result, this shows that
Eq.~\eqref{eq:K2X} has exactly one admissible solution.
It follows that $A=1-X$ is uniquely determined, and then
\begin{eqnarray}
C
=
\frac{1-\mathcal{K}_1X}{1-X}
\end{eqnarray}
is unique as well.
Since $a>0$ and $0<i\leq\pi/2$, the values
$A=a^2$ and $C=\cos^2 i$ uniquely recover
$a=\sqrt{A}$ and $i=\arccos\sqrt{C}$.
Therefore
\begin{eqnarray}
(a,i)
=
(a',i').
\end{eqnarray}

With the parameters now known to be equal,
Eq.~\eqref{eq:rigidprop} says that the two transformed copies of the
same polynomial have the same zero set.
It follows that the relative rigid motion $E'^{-1}\circ E$ maps the
full algebraic curve $V(P)$ onto itself.
Because both $E$ and $E'$ are orientation preserving, so is this
relative rigid motion.
As shown in Sec.~\ref{sec:leading}, this curve has no nontrivial
orientation-preserving rigid-motion symmetry.
Consequently,
\begin{eqnarray}
E'^{-1}\circ E
=
\mathrm{id},
\end{eqnarray}
and hence $E=E'$.
This completes the global uniqueness argument.

As an immediate consequence, the full critical curve also determines
$(a,i)$ uniquely and globally, even when its position and orientation
on the screen are unknown.
More precisely, if
$E\!\left(C(a,i)\right)=E'\!\left(C(a',i')\right)$ for two parameter
pairs in the domain \eqref{eq:domain} and two orientation-preserving
rigid motions $E$ and $E'$, then
$(a,i)=(a',i')$ and $E=E'$.
Indeed, restricting the equality of the full curves to any
positive-length segment gives the hypothesis of the result proved
above.

The limiting values $a=0$ and $a=1$ behave differently from the
rotating non-extremal case.
At $a=0$, Eqs.~\eqref{eq:spoell} and \eqref{eq:spoq} are singular as
written, so the Schwarzschild critical curve is more conveniently
described directly by
\begin{eqnarray}
\alpha^2+\beta^2=27 .
\label{eq:schwcurve}
\end{eqnarray}
It is therefore a circle of radius $3\sqrt{3}$ centered at the origin
~\cite{Synge:1966okc}.
Consequently, every segment is a circular arc, and its geometry is
independent of $i$.
A positive-length circular arc determines its supporting circle.
Moreover, such an arc cannot coincide under a rigid motion with a
segment of a rotating critical curve with $0<a<1$.
If it did, Eq.~\eqref{eq:idealgen} would imply that the irreducible
degree-six polynomial $P$ is proportional to the degree-two polynomial
defining the circle, which is impossible.
Thus $a=0$ can be distinguished from the rotating case.
The inclination $i$, however, is not determined, because the
Schwarzschild critical curve is independent of $i$.
In addition, the rotational symmetry of the circle makes a rotation of
the screen unobservable.

At the other limiting value, $a=1$, the critical curve has a different
near-horizon structure.
In the extremal limit, it contains a positive-length near-horizon
extreme Kerr line for a range of inclinations
~\cite{Gralla:2017ufe}.
For any two distinct inclinations in this range, equal-length
subsegments can be chosen from their near-horizon lines, and these
straight segments are congruent under an orientation-preserving rigid
motion of the screen.
Thus an arbitrary positive-length segment does not determine the
inclination uniquely in the extremal case.

\section{Conclusion}
\label{sec:conclusion}
We have established a local-to-global uniqueness result for segments of
the critical curve of a rotating non-extremal Kerr black hole.
Any connected positive-length segment determines the dimensionless spin
parameter $a$ and the inclination angle $i$ uniquely over the full
domain $0<a<1$ and $0<i\leq\pi/2$, even when the position and
orientation of the segment on the observer's screen are unknown.
More precisely, if two such segments coincide as point sets after
orientation-preserving rigid motions of the screen, then both their
parameter pairs and the corresponding rigid motions are identical.

The proof is based on an algebraic representation of the critical
curve.
Eliminating the spherical photon-orbit radius gives an irreducible
degree-six polynomial $P$.
The vanishing ideal of a positive-length segment is generated by the
corresponding rigid-motion transform $P\circ E^{-1}$.
Consequently, the segment determines the entire supporting algebraic
curve, while this irreducible generator is determined uniquely up to
multiplication by a nonzero real constant.
The leading form of $P$ identifies the intrinsic directions of the
curve, while its unique reflection axis and the absence of nontrivial
orientation-preserving rigid-motion symmetries ensure that its position
and orientation on the screen are uniquely determined.
Finally, two rigid-motion invariants determine $a^2$ and $\cos^2 i$
uniquely throughout the admissible parameter domain.

The limiting spin cases require separate treatment.
For $a=0$, the critical curve is a circle independent of the inclination
angle.
The Schwarzschild case can therefore be distinguished from the rotating
case, but neither the inclination angle nor the orientation of the
critical curve on the screen is uniquely determined.
The extremal case $a=1$ is not covered by the present result.
For a range of inclinations, the extremal critical curve contains a
positive-length near-horizon line.
Equal-length subsegments of this line at different inclinations are
congruent under orientation-preserving rigid motions, so an arbitrary
positive-length segment does not determine the inclination uniquely.

The result obtained here concerns exact geometric identifiability and
assumes that a connected positive-length segment is known exactly as a
point set.
A complementary numerical framework defined geometric observables for
standardized segments of the critical curve and examined the
sensitivity of the reconstructed parameters to artificial Gaussian
perturbations in those observables~\cite{Hioki:2026xch}.
The present uniqueness result supplies the underlying geometric
identifiability, while that perturbation test diagnoses how
uncertainties in the observables propagate through the
parameter-recovery procedure.
The numerical test does not model the full observational pipeline,
including image reconstruction, segment identification, finite
sampling, and the estimation of the observables.
A full treatment of these observational uncertainties remains an
important direction for future work.

A separate theoretical direction is to investigate whether the
algebraic strategy developed here extends to charged or more general
rotating black hole spacetimes.

\section*{Acknowledgements}
We would like to thank Umpei Miyamoto for useful discussions.



\end{document}